# Doing AI: Algorithmic decision support as a human activity

Joachim Meyer

Dept. of Industrial Engineering

Tel Aviv University

**Author Note**

Joachim Meyer   https://orcid.org/0000-0002-1801-9987

## Abstract

Algorithmic decision support (ADS), using Machine-Learning-based AI, is becoming a major part of many processes. Organizations introduce ADS to improve decision-making and use available data, thereby possibly limiting deviations from the normative "homo economicus" and the biases that characterize human decision-making. However, a closer look at the development and use of ADS systems in organizational settings reveals that they necessarily involve a series of largely unspecified human decisions. They begin with deliberations for which decisions to use ADS, continue with choices while developing and deploying the ADS, and end with decisions on how to use the ADS output in an organization's operations. The paper presents an overview of these decisions and some relevant behavioral phenomena. It points out directions for further research, which is essential for correctly assessing the processes and their vulnerabilities. Understanding these behavioral aspects is important for successfully implementing ADS in organizations.

## Introduction

Algorithms have been entering our lives for quite some time, and the recent rapid development of Generative Artificial Intelligence (GenAI) and Large Language Models has given rise to enthusiasm and great expectations but also concerns. One important application area for Artificial Intelligence (AI) and machine learning (ML) is algorithmic decision support (ADS), intended to support human decision-makers or even to replace them in some processes. Organizations can deploy these systems, like previous types of decision support, to improve decision-making effectiveness (Clark Jr., Jones, & Armstrong, 2007). Their development and integration into an organization's operations require a substantial commitment of resources, which is only justified if the organization expects clear benefits from using the system.

Using algorithms to support decisions may imply that human biases, fallacies, and limitations cease to affect decisions when ADS is part of the process. However, the current paper aims to demonstrate that the development and deployment of ADS involves human decision-making at numerous points with all its quirks. The behavioral decision-making community has an essential role in the identification and study of the properties of human decisions that are relevant for ADS and in developing tools to limit possible adverse effects.

The paper will briefly describe the uses of ADS and perceptions regarding its objectivity. It will then discuss ADS as a human activity and an organizational process. It then describes three types of decisions related to ADS – the decision whether to develop ADS to support a process, the decisions made during the development and deployment of ADS, and decisions regarding the output from ADS. Some conclusions are drawn regarding the important contributions of behavioral decision-making research to the successful use of ADS.

## Some ADS achievements and reservations

ADS has been studied in various contexts. For instance, in legal contexts, algorithms can support bail decisions with possible large reductions in crime or jailing rates (Kleinberg, Lakkaraju, Leskovec, Ludwig, & Mullainathan, 2018). Similarly, ADS can help make decisions regarding child welfare (Saxena, Badillo-Urquiola, Wisniewski, & Guha, 2020). There has been a trend toward "personalized evidence-based medicine," based on analyzing electronic medical records with algorithmic tools (Kent, Steyerberg, & van Klaveren, 2018). AI-based analyses can lead to better medical diagnostic or therapeutic decisions (Puaschunder, Mantl, & Plank, 2020), such as the early diagnosis of breast cancer (McKinney et al., 2020). Even when AI does not improve diagnostic decisions, it may lower the workload of the clinician who analyzes, for instance, mammographs (Lång et al., 2023). In some areas, such as financial markets, algorithmic advisors provide clear benefits (Tao, Su, Xiao, Dai, & Khalid, 2021), and they are the basis for investment decisions in high-frequency, algorithmic trading (Virgilio, 2019).

Human decision-making differs in many cases from the prescriptions of normative decision theory (Takemura, 2021). In contrast, algorithmic decision support may be seen as a mathematical, systematic way to make decisions, eliminating human fallibility. As such, it continues a tradition of searching for methods to create objective, rational decision-making processes that lead to better decisions. Already in the 1960s, there was much optimism about the potential of using operations research and management science to make decisions with scientific methods, perhaps replacing human decision-makers and their limitations (Dando & Bennett, 1981). A crisis in the 1980$^{th}$ followed this enthusiasm, with practitioners showing diminishing interest in mathematical methods (Ackoff, 1987; Corbett & Van Wassenhove, 1993).

Similarly, in the early days of data science, there was optimism that data science could replace scientific theories (Anderson, 2008), and big data was seen as possibly leading to a "management revolution" (McAfee & Brynjolfsson, 2012). In the meantime, this optimism has become somewhat dampened, partly due to the realization that together with the potential benefits from ADS, there are also various ethical, legal, and social concerns (Čartolovni, Tomičić, & Lazić Mosler, 2022), partly related to possible biases due to the choice of the data or the algorithmic processes (Beil, Proft, van Heerden, Sviri, & van Heerden, 2019). Also, introducing ADS can adversely affect the system and create phenomena such as "flash crashes" or similar anomalies in financial markets (Min & Borch, 2022). One should clearly avoid "big data hubris," the belief that big data can replace traditional analytical methods (Lazer, Kennedy, King, & Vespignani, 2014). Still, despite the growing awareness of possible limitations, AI-based ADS are developed for industry and business (Gupta, Modgil, Bhattacharyya, & Bose, 2021), medicine (Antoniadi et al., 2021), the public sector (Kuziemski & Misuraca, 2020), and other fields.

## ADS as a human activity

ADS may seem to be a mechanistic process in which data is fed into a machine (the algorithm), which provides necessarily correct output that can be trusted because it is based on mathematics. However, a closer look at ADS reveals that it requires a sequence of closely intertwined human activities that are related to three aspects of the process – the decision of whether to use AI for a decision, the process of generating AI output, and the decision of how to use the output for a given purpose (see Figure 1). The three aspects of developing and using ADS are not separate. For instance, the plans on how to use the ADS output will be part of the deliberation on whether to develop ADS for a specific decision. Similarly, the quality of the AI output, which may depend on the process used to generate it, will be important when choosing how to utilize it.

So far, we still lack an understanding of these activities and their determinants, although there is awareness that they can be subject to various biases (Arnott, 2006; Richardson, 2022). In the following, I will briefly discuss the three aspects, arguing that the process leading to ADS is an important topic for research on behavioral decision-making. The results of this research can provide valuable input for ADS developers, organizations that employ it, and the public that is served or experiences the results of the supported decision processes.

Figure 1: The three aspects of developing algorithmic decision support – whether to do it, how to do it, and what to do with it.

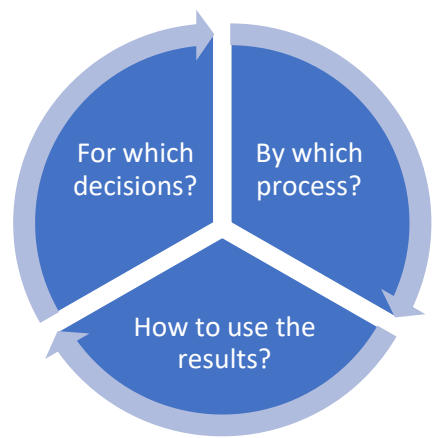

## ADS as an organizational process

The abundance of research on data-based AI and ML may obscure the fact that implementing an ADS and making it part of an organization's operations requires the continuing investment of considerable resources. The deployment of ADS results from a systematic process in which the support was specified, its technical aspects were implemented, they were tested, parameters were tuned, and eventually, it was launched in some organizational setting. While it is in use, someone maintains it, pays for it, and may (or should) evaluate it constantly, updating it when necessary or when opportunities arise. Thus, deploying an ADS system is preceded and accompanied by a series of organizational decisions that determine the system specification and design, the computations needed to generate its output, and the use of the output for actual decisions. The entire process depends on various organizational factors, as well as human decisions and involvement (Golightly, Kefalidou, & Sharples, 2018).

The introduction of ADS into an organization is not just a technical issue. Rather, it may require cultural adjustments in the organization (Fountaine, McCarthy, & Saleh, 2019). This poses major challenges to management, which needs to address the algorithms' rapidly changing abilities and the requirements and constraints that arise when AI is introduced into organizational processes (Berente, Gu, Recker, & Santhanam, 2021).

The rapid developments in AI and ML and the realization of their possible application for decision support led to an immense increase in related research. For instance, the number of publications with "Decision Support" in any field in Clarivate Web of Science doubled from 24,063 in 2014 to 48,684 in 2021. There are, however, relatively few studies of the actual implementation of decision support in organizations. One review on hospital-based computerized decision support systems identified 1836 papers that dealt with the topic between 2010 and 2021. Only 9 dealt with the costs of implementing such a system (Donovan, Abell, Fernando, McPhail, & Carter, 2023). Indeed, despite the abundance of research on ADS, studies on the actual implementations of ML-based decision support in clinical settings (or other domains) are still few (Susanto, Lyell, Widyantoro, Berkovsky, & Magrabi, 2023). Similarly, a review of the economic impacts of decision-support systems in hospitals found that out of 3113 potentially relevant articles published between 2010 and 2023, only 29 dealt with the economic consequences of using such a system (White et al., 2023). Also, practitioners' acceptance of decision support is not universal, and even if a system has been developed and deployed, it may not be used. A review of studies on the use of decision support in clinical settings showed an average uptake in only 34.2% of events where the system could be used (Kouri, Yamada, Lam Shin Cheung, Van de Velde, & Gupta, 2022). For most events, practitioners chose not to use decision support. These studies show that we know relatively little about the conditions for successful use of ADS in organizations.

## Can or should ADS be developed?

The ADS process begins with the question of whether decision support should be developed. This complex issue relates to operational, regulatory, organizational, psychological, or economic considerations. One way to map them is to consider decision support as a type of automation, similar to the definition that automation is "a device or system that accomplishes (partially or fully) a function that was previously, or conceivably could be, carried out (partially or fully) by a human operator" (Parasuraman, Sheridan, & Wickens, 2000; p. 287). In human-automation interaction, the issue of function allocation is critical (Challenger, Clegg, & Shepherd, 2013). It is the decision of how to implement and assign system functions for them to be performed by humans, by equipment, or by both (Beevis, 1992). Four rules for function allocation have been proposed (Price, 1985). They can serve as a conceptual framework for analyzing whether to develop an ADS in a given setting.

*Mandatory allocation* assigns tasks according to laws or regulations that determine what should be done by humans or automation. For ADS, mandatory restrictions exist in many areas. For instance, Article 22 in the E.U. General Data Protection Regulation (GDPR) states that a person has the right not to be subject to algorithmic decisions on important aspects of their life (Roig, 2017), limiting the decisions that can be automated and the possible designs for ADS. Similarly, the E.U. AI Act from 2021, which is intended to regulate the use of AI in Europe (Veale & Zuiderveen Borgesius, 2021), classifies risks and forbids the use of AI for certain purposes, such as manipulating people in ways that may cause harm. High-risk systems, such as systems that manage critical infrastructure, employees, services, or benefits, are regulated, must be registered, and must have human oversight, limiting the use of ADS for decisions in these areas, similar to the requirements in the GDPR. The demand for human involvement is particularly pertinent when systems, most notably autonomous weapon systems, are built to harm some people. For them, there is an explicit demand for "meaningful human control" and human involvement in decisions that can hurt people (Boulanin, Davison, Goussac, & Peldán Carlsson, 2020).

*Balance of value allocation* assigns tasks according to the quality of the outcome that can be achieved when a human or automation performs an action ("the one who can do it best should do the task"). For some decisions, computers are indeed better. For instance, the human ability to control highly complex dynamic systems, such as industrial processes, is limited, and their functioning depends on algorithmic decisions (Ng, Chen, Lee, Jiao, & Yang, 2021). When either AI or humans can make decisions, the relative quality of the decisions often depends on the human's expertise. A fairly consistent finding in research on aided decision-making is that the combination of a human and an aid is not better than the better of the two alone (Meyer & Kuchar, 2021). Providing less knowledgeable users with decision support can lead to worse results than automating the decision, while expert users benefit little from the support (Inkpen et al., 2023). Thus, organizations may have to consider the differential use of ADS, depending on the characteristics of the people involved in the process.

*Utilitarian and cost-based allocation* determines function allocation based on cost-benefit analyses, considering operational costs, efficiency, or the values of outcomes. This point is related to questions of the efficiency of decision processes, which may have to be balanced against human involvement and perceived fairness (Zarsky, 2016). How efficient should an AI decision process be to be acceptable for making decisions? To what extent should one sacrifice efficiency to involve a human in decisions? For instance, an algorithm processing thousands of applications per hour can easily make credit decisions. A human, in contrast, needs several minutes to deliberate carefully about a credit application; therefore, a person handles orders of magnitude fewer applications than the algorithm. Are we willing to accept a slowing of the processing of applications to ensure that a person views each application? What criteria should be used to determine whether a reduction in efficiency is justified to have human involvement?

*Allocation of functions for affective or cognitive support* considers the affective or cognitive needs of the humans involved in the process. The decision whether to use ADS is related to the views held by the public or stakeholders regarding the value of algorithmic support. This is related to phenomena such as algorithm aversion, where people tend to avoid algorithmic decision support compared to their response to similar support from people (Dietvorst, Simmons, & Massey, 2015). However, this is not a general phenomenon, and there may also be a preference for algorithmic judgments (Logg, Minson, & Moore, 2019). The attitude towards ADS may be somewhat akin to evaluations of the safety benefits of autonomous driving. Autonomous vehicles may be less likely to be involved in accidents than vehicles driven by humans (Hicks, 2018). Still, the tolerance for accidents involving an autonomous vehicle is lower than that for accidents involving a human driver, imposing much stricter safety requirements on the autonomous vehicle (Liu, Du, & Xu, 2019).

The decision to develop an ADS will be based on a combination of considerations related to the four rules. People involved in the process (in the organization or outside service providers) will have to decide on implementing the different rules and the relative weight given to each of them. This will determine whether a system should be developed and how it will be designed. The choices will depend, to a large extent, on perceptions of the problems, evaluations of outcomes, and other factors considered in behavioral decision-making.

## ADS development as a series of decisions

The human involvement in the process does not end once an organization deploys an ADS for specific tasks. The development and use of the ADS involves a series of actions and decisions, depicted schematically in Figure 2. It begins with identifying the data sources that can be used as input for the algorithm. The choice of which sources to use will depend on the properties of the data and especially its quality, which is not only the simple accuracy of data but also numerous other properties, such as the availability of the data, its being up-to-date, complete, without duplicates, or unbiased (Sidi et al., 2012). Data never fully represents reality since only certain parts of it are recorded. Systematic differences between what exists in a data set and the reality it supposedly reflects must be considered when using data as the basis for ADS (Meyer, 2024). In particular, concerns about possible biases in the data must be addressed. These depend on the data set on which the algorithm was trained and other biasing factors (Mehrabi, Morstatter, Saxena, Lerman, & Galstyan, 2021). However, it is difficult to predict what will bias results. For example, excluding race as a predictor from algorithms for determining college admissions might seem reasonable. Still, the inclusion of race can actually increase the admission process's equity and efficiency (Kleinberg, Ludwig, Mullainathan, & Rambachan, 2018).

Figure 2: The algorithmic analytical process as a series of human decisions and actions (*indicated in italics*)

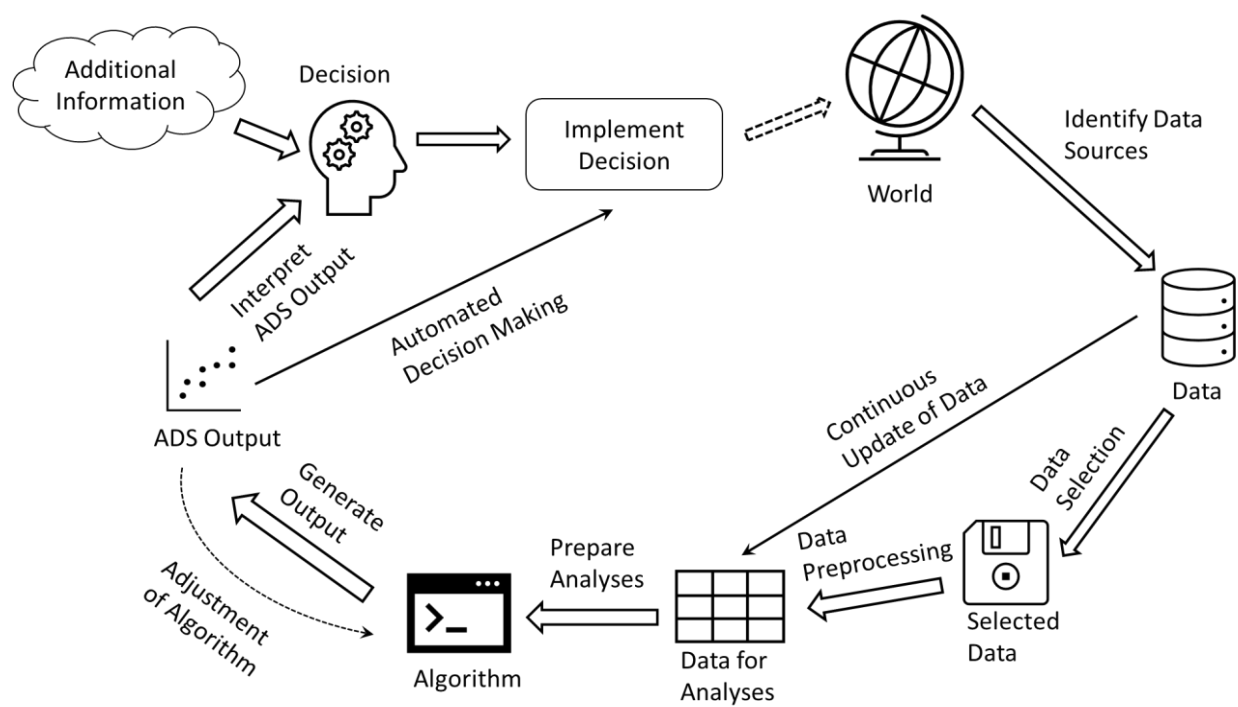

After choosing data sources that serve as the basis for the ADS, the specific data for the analysis needs to be selected and preprocessed. This requires determining the time frame for which data will be included (the last 3 months, year, 3 years, 10 years, etc.), the geographic areas from which the data will be analyzed (an entire country, certain parts, without rural areas, without certain areas with specific characteristics, etc.), whether to exclude certain times (holidays, weekends, summer, Covid-19 restrictions, etc.) and what to define as anomalies or outliers that should be excluded from the data set. Also, data may be adjusted to facilitate processing by combining, for instance, occupations into broader categories. The data selection and preprocessing involve various decisions the analysts need to make without necessarily evaluating each choice's implications on the resulting data set and the subsequent outcomes of the analyses.

The process is complicated by the fact that most data sources are dynamic. The incoming data changes due to changes in the measured phenomena or in the recording of events. This will affect the ability of the algorithm to predict events. For instance, the Google Flu Trends (GFT) model initially predicted the occurrence of seasonal flu from search terms quite well. After some time, case predictions were greatly overestimated and had only limited value (Lazer et al., 2014). Because of the possibility of such changes, the ADS must be reevaluated, and procedures in the organization using the ADS may need to be adjusted if the quality of its predictions changes.

Once the data set is ready, one needs to determine the method by which the algorithmic output is generated. The same data set can be analyzed in various ways, such as using different algorithms or setting different values of meta-parameters for the algorithms. Differences in the analytical process can lead to different results. For instance, twenty-nine groups received the same data to determine whether soccer players with darker skin color receive more red cards (are removed from the game) than players with lighter skin (Silberzahn et al., 2018). The groups analyzed the results with 21 unique covariate combinations, and about two-thirds of the groups showed a significant effect in the expected direction, while one-third did not. Similarly, in an extensive study on the analysis of medical resonance imaging (MRI) data, 70 groups analyzed the same data. The workflow of the analytics process was different for all groups, leading to very different results of the analyses (Botvinik-Nezer et al., 2020). Also, different research groups used different processes for analyzing the same data set for a microeconomic causal model, reporting different results (Huntington-Klein et al., 2021). Regarding scientific activities, different analysts used different approaches to test hypotheses on the participation of male or female scientists in an online forum, reaching different results (Schweinsberg et al., 2021). Thus, there can be great variability in the results obtained from analyzing the same data set to answer the same questions.

The choices during the analytics process are often not made explicit despite their possible impact on the results. Also, the process is not simply an application of different tools to the same data set. Analysts may explore different meta-parameters or choices regarding the data. A concern in this context is that they may approach the problem with preconceptions, and analyses may continue until they provide results that align with expectations. This process may be subject to various biases affecting perceptions of fairness and the acceptance of the algorithm results (Kordzadeh & Ghasemaghaei, 2022).

The results can be presented in many ways, each making certain aspects more or less salient (Eisler & Meyer, 2023). Here, too, exists the potential for biases in the process (Akter et al., 2021). The analysts may choose depictions that correspond to their preconceived notions of the desired results (after all, much effort has gone into the process, and the analysts usually want it to succeed), manifesting the confirmation bias in the interpretation of the results of the analysis (Lehner, Adelman, Cheikes, & Brown, 2008).

## Using ADS output

The considerations of how to use the ADS output are similar to those during the decision to develop the system. Here, issues of utility, mandatory requirements, and cognitive and affective effects of having the system also will be relevant.

The choice to introduce ADS into a decision process is most likely motivated by the hope of achieving improved decisions, either in terms of the quality of the decision or the cost of making it. The quality of the ADS for computer science or statistics researchers is usually defined by its ability to provide an indication that corresponds to some objectively known "ground truth." It is quantified with measures of this correspondence, such as precision and recall, the area under the curve (AUC), the F1 score, etc. (Padilla, Netto, & da Silva, 2020). However, ADS's value depends on its benefits to the organization and stakeholders. This assessment goes beyond the measurement of statistical properties of the algorithmic output. It needs to consider the extent to which decisions with the ADS are better than those without it and how important the improvement in the decision quality is for the organization. Assessing these values is often complicated because of the limited availability of data for evaluating the actual implications of ADS use and the complexity of measuring the impact of ADS in actual economic settings (Kleinberg, Lakkaraju, et al., 2018).

The effects of the ADS on the quality of the decisions will depend on how it is integrated into the decision procedures. ADS results can be used directly in some automatic decision process, perhaps supervised by a human whose task is to intervene if abnormal events occur. Such automated decision-making is already widely used, for instance, in controlling complex networks or algorithmic trading in online markets. Alternatively, the algorithm output can be information for a human decision-maker. She must have access to additional information besides the ADS output to be able to make any meaningful contribution to the decision. The person needs to combine the information from the ADS with the additional information and decide, based on all available information. A major challenge in this context is the weight given to the information from the different sources. Assigning incorrect weights to the information from the ADS can impair decision quality, compared to automating the decision (relying entirely on the ADS) or letting the human decide without the ADS (Meyer & Kuchar, 2021). This is a valid concern, considering that empirical research has shown that decision aids often receive insufficient but also sometimes excessive weight in decision-making (Bartlett & McCarley, 2019; Meyer & Lee, 2013; Meyer, 2001; Parasuraman & Riley, 1997).

Introducing ADS into a decision process may cause major changes that must be considered when implementing such a system. For one, people will not simply continue to do what they have done so far, only now with the support of ADS. Rather, introducing automation (such as ADS) most likely changes the behavior of those interacting with the system (Dekker & Woods, 2002; Roth, Sushereba, Militello, Diiulio, & Ernst, 2019). For instance, automation bias may cause the user to rely strongly on the automation, i.e., the ADS output, without analyzing the available information (Goddard, Roudsari, & Wyatt, 2012, 2014; Lyell & Coiera, 2017; Mosier, Skitka, Heers, & Burdick, 1997; Wickens, Clegg, Vieane, & Sebok, 2015). This issue has been cause for particular concern in medical systems, where the introduction of diagnostic aids can lead to a strong reliance of physicians on the aid output without them independently evaluating the patient's data (Goddard et al., 2012, 2014; Lyell & Coiera, 2017; Mosier et al., 1997; Wickens et al., 2015). In such cases, the medical staff's skills may be degraded because indications from the ADS primarily trigger behavior, and no independent decisions are made (Sutton et al., 2020).

The mandatory requirements for using an ADS in a decision process create various dilemmas and raise questions. What does it mean to have a human involved in a process? Is a credit decision made by a human bank employee, based on the output of an ADS, without the human having independent information about the applicant, a decision that a human makes? What makes such a decision acceptable, as far as the requirements of GDPR go? Few systematic methods to quantify human involvement exist. One attempt to create such a measure is the ResQu (responsibility quantification) model, which analyzes human involvement and responsibility for outcomes in decisions with intelligent support (Douer & Meyer, 2020). It uses Information Theory to quantify human involvement in a process, defining a person's responsibility as the proportion of the entropy reduction due to the person. It is also a fairly good descriptive model for predicting the impression people have about the relative importance of a human in an aided decision process (Douer & Meyer, 2021). The model predicts that human responsibility will rapidly diminish with automation that performs a classification task better than a human. For automation to be deployed, it must perform the task as well as a human, and probably better. Thus, one can expect that human responsibility will be limited when ADS is used. Analytical tools, similar to the ResQu model, need to be developed to provide systematic frameworks for evaluating the implications of using ADS in a given setting.

The ADS may have complex cognitive or affective implications for people interacting with it or participating in supported processes. For example, even if an ADS functions well, there may still be a requirement to keep a human involved in the decisions. This person may serve as a "moral crumple zone" (Elish, 2019). The crumple zone in a car absorbs some of the energy of a collision, thereby limiting the harm to the passengers. Similarly, the person involved in the ADS-supported process may be used to protect the ADS developers and deployer from carrying the moral and legal responsibility for system failures. Also, in a largely well-functioning system with rare failures, the person will most likely approve almost all algorithmic decisions. Contradicting the ADS, possibly introducing a failure that would not have happened if the person had adhered to the ADS advice, will very likely have negative consequences for the person. Thus, the person may face a no-win situation where either overriding or following the automation indication can have negative results. We still know little about people's decisions when they find themselves in such roles.

## Conclusions

Most work on ADS has aimed so far to demonstrate that it is possible. Now, the technology is sufficiently developed to be implemented in actual organizational settings. This change gives rise to many practical questions related to the planning of the ADS, its development and maintenance, and the use of its output. A closer inspection of the ADS process shows that it is a human activity, subject to the biases and limitations of human decision processes. However, it can also benefit from human strengths like flexibility and creativity.

The development of ML-based decision support touches on various behavioral decision-making phenomena at numerous points in the process. These include status quo and default biases in choosing data, preprocessing steps, algorithms, presentations, or interpretations. Also, confirmation bias may affect the evaluation of outcomes and the steps taken to alter the outcome when trying to improve it. A larger-scale review evaluated the cognitive biases in the context of rule-based decision support (Kliegr, Bahník, & Fürnkranz, 2021). These biases remain relevant for systems using machine learning and explainable AI (Bertrand, Belloum, Eagan, & Maxwell, 2022). Similarly, decision-makers or analysts' preferences, attitudes, mental models, or beliefs may affect the processes or their outcomes, possibly leading to systematic biases or non-optimal use of available information for decision-making. The behavioral decision-making community can contribute much to identifying such phenomena and mitigating their negative effects. Many additional issues need to be considered, including the changes in the selection and training of employees when an ADS is introduced or the design of incentive systems for employees working with ADS.

One may ask whether the issues described here are new and arose due to the use of ML and AI or whether they exist in any decision support. Indeed, older forms of decision support that did not rely on AI involved numerous decisions along the way that resemble those presented here (Power, 2008). Older knowledge-based decision support systems consisted of sets of rules elicited from experts. They also required extensive development, maintenance to keep the rule system up-to-date, and organizational commitment that specified how the system should be used. However, the newer AI-based "non-knowledge based" systems pose new problems (Sutton et al., 2020). ML algorithms, particularly "deep learning, " generate results that can often not be intuitively understood. One needs to accept them based on trust in the system. Some support for this trust can be generated by "explainable AI" (XAI), which provides explanations for the reasons why the output was generated (Antoniadi et al., 2021). However, the effect of explanations may be limited and may depend on the user's expertise (Bayer, Gimpel, & Markgraf, 2022).

One conclusion from the points raised here is that an ADS is often more than a simple tool added to a procedure to improve decision-making. The introduction of ADS should be accompanied by the development of tools that can help people build an ADS system. For instance, one precondition for understanding alternatives and evaluating the implications of certain decisions is knowing what would happen if different choices were made in the process. Relatively few tools exist for such analyses. In most cases, the analyst must rerun the analyses, change the input data, adjust meta-parameters, or use different analytical tools. One example of such a tool is the "what-if" tool (Wexler et al., 2020), which allows the interactive exploration of ML models, including the exploration of data features, changes in the input data, evaluation of different models, etc.

So far, we still lack a general model for predicting the attitudes towards algorithmic support. To be useful, such a model must consider variables related to the task that is supported, the quality of the algorithmic support, the ability to make adequate decisions without the support, the conditions in which decisions are made (time pressure, social settings, etc.), and the individual characteristics of the decision maker. A comprehensive empirical investigation of all possible combinations of the relevant variables is impossible, so more abstract models need to be developed to predict different people's acceptance of algorithmic support in specific settings. The development of such models can possibly be one of the major challenges for behavioral decision-making research in the coming years.

The development of ADS and its introduction into decision-making give rise to numerous issues related to human decision-making. The behavioral decision-making community has an important role in understanding human decisions during the development or use of an ADS. The work can expand conceptual frameworks and have definite value for ADS's development, deployment, and use. For the research to be meaningful, it must breach disciplinary boundaries, combining decision-theoretic and mathematical tools, operations management, behavioral economics, social psychology, and ethnographic methods for studying the actual practice of using

ADS in an organization and the changes accompanying its introduction and use. The questions that arise in the context of the development of ADS can open new directions for empirical research and theoretical developments and can change and invigorate existing ones. It is very well possible that this is a starting point for a new form of decision research for the age of intelligent systems.

Bibliography

Ackoff, R. L. (1987). Presidents' symposium: OR, A post mortem. Operations Research, 35(3), 471–474. doi: 10.1287/opre.35.3.471

Akter, S., McCarthy, G., Sajib, S., Michael, K., Dwivedi, Y. K., D'Ambra, J., & Shen, K. N. (2021). Algorithmic bias in data-driven innovation in the age of AI. International Journal of Information Management, 60, 102387. doi: 10.1016/j.ijinfomgt.2021.102387

Anderson, C. (2008). The end of theory: The data deluge makes the scientific method obsolete. *Wired Magazine*, *16*(7), 16–07.

Antoniadi, A. M., Du, Y., Guendouz, Y., Wei, L., Mazo, C., Becker, B. A., & Mooney, C. (2021). Current Challenges and Future Opportunities for XAI in Machine Learning-Based Clinical Decision Support Systems: A Systematic Review. *Applied Sciences*, *11*(11), 5088. doi: 10.3390/app11115088

Arnott, D. (2006). Cognitive biases and decision support systems development: a design science approach. *Information Systems Journal*, *16*(1), 55–78. doi: 10.1111/j.1365-2575.2006.00208.x

Bartlett, M. L., & McCarley, J. S. (2019). Human interaction with automated aids: Implications for robo-advisors. *Financial Planning Review*, *2*(3–4). doi: 10.1002/cfp2.1059

Bayer, S., Gimpel, H., & Markgraf, M. (2022). The role of domain expertise in trusting and following explainable AI decision support systems. *Journal of Decision Systems*, *32*(1), 110–138. doi: 10.1080/12460125.2021.1958505

Beil, M., Proft, I., van Heerden, D., Sviri, S., & van Heerden, P. V. (2019). Ethical considerations about artificial intelligence for prognostication in intensive care. *Intensive Care Medicine Experimental*, *7*(1), 70. doi: 10.1186/s40635-019-0286-6

Berente, N., Gu, B., Recker, J., & Santhanam, R. (2021). Managing artificial intelligence. *MIS Quarterly*, *45*(3), 1433–1450.

Bertrand, A., Belloum, R., Eagan, J. R., & Maxwell, W. (2022). How Cognitive Biases Affect XAI-assisted Decision-making: A Systematic Review. *Proceedings of the 2022 AAAI/ACM Conference on AI, Ethics, and Society*, 78–91. New York, NY, USA: ACM. doi: 10.1145/3514094.3534164

Botvinik-Nezer, R., Holzmeister, F., Camerer, C. F., Dreber, A., Huber, J., Johannesson, M., … Schonberg, T. (2020). Variability in the analysis of a single neuroimaging dataset by many teams. *Nature*, *582*(7810), 84–88. doi: 10.1038/s41586-020-2314-9

Boulanin, V., Davison, N., Goussac, N., & Peldán Carlsson, M. (2020). *Limits on Autonomy in Weapon Systems: Identifying Practical Elements of Human Control*. SIPRI: Stockholm International Peace Research Institute.

Čartolovni, A., Tomičić, A., & Lazić Mosler, E. (2022). Ethical, legal, and social considerations of AI-based medical decision-support tools: A scoping review. *International Journal of Medical Informatics*, *161*, 104738. doi: 10.1016/j.ijmedinf.2022.104738

Challenger, R., Clegg, C. W., & Shepherd, C. (2013). Function allocation in complex systems: reframing an old problem. *Ergonomics*, *56*(7), 1051–1069. doi: 10.1080/00140139.2013.790482

Clark Jr., T. D., Jones, M. C., & Armstrong, C. P. (2007). The dynamic structure of management support systems: theory development, research focus, and direction. *MIS Quarterly*, *31*(3), 579–615.

Corbett, C. J., & Van Wassenhove, L. N. (1993). The natural drift: what happened to operations research? *Operations Research*, *41*(4), 625–640. doi: 10.1287/opre.41.4.625

Dando, M. R., & Bennett, P. G. (1981). A Kuhnian crisis in management science? The Journal of the Operational Research Society, 32(2), 91. doi: 10.2307/2581256

Dekker, S. W. A., & Woods, D. D. (2002). MABA-MABA or Abracadabra? Progress on Human-Automation Co-ordination. *Cognition, Technology & Work*, *4*(4), 240–244. doi: 10.1007/s101110200022

Dietvorst, B. J., Simmons, J. P., & Massey, C. (2015). Algorithm aversion: people erroneously avoid algorithms after seeing them err. *Journal of Experimental Psychology: General*, *144*(1), 114–126. doi: 10.1037/xge0000033

Donovan, T., Abell, B., Fernando, M., McPhail, S. M., & Carter, H. E. (2023). Implementation costs of hospital-based computerised decision support systems: a systematic review. *Implementation Science*, *18*(1), 7. doi: 10.1186/s13012-023-01261-8

Douer, N., & Meyer, J. (2020). The responsibility quantification model of human interaction with automation. *IEEE Transactions on Automation Science and Engineering*, *17*(2), 1044–1060. doi: 10.1109/TASE.2020.2965466

Douer, N., & Meyer, J. (2021). Theoretical, measured, and subjective responsibility in aided decision making. *ACM Transactions on Interactive Intelligent Systems*, *11*(1), 1–37. doi: 10.1145/3425732

Eisler, S., & Meyer, J. (2023). Visual analytics and human involvement in machine learning. In L. Rokach, O. Maimon, & E. Shmueli (Eds.), *Machine learning for data science handbook: data mining and knowledge discovery handbook* (pp. 945–970). Cham: Springer International Publishing. doi: 10.1007/978-3-031-24628-9_40

Elish, M. C. (2019). Moral Crumple Zones: Cautionary Tales in Human-Robot Interaction. *Engaging Science, Technology, and Society*, *5*, 40. doi: 10.17351/ests2019.260

Fountaine, T., McCarthy, B., & Saleh, T. (2019). Building the AI-Powered Organization. *Harvard Business Review*, *2019*(July-August).

Meyer

21Goddard, K., Roudsari, A., & Wyatt, J. C. (2012). Automation bias: a systematic review of frequency, effect mediators, and mitigators. *Journal of the American Medical Informatics Association*, *19*(1), 121–127. doi: 10.1136/amiajnl-2011-000089

Goddard, K., Roudsari, A., & Wyatt, J. C. (2014). Automation bias: empirical results assessing influencing factors. *International Journal of Medical Informatics*, *83*(5), 368–375. doi: 10.1016/j.ijmedinf.2014.01.001

Golightly, D., Kefalidou, G., & Sharples, S. (2018). A cross-sector analysis of human and organisational factors in the deployment of data-driven predictive maintenance. *Information Systems and E-Business Management*, *16*(3), 627–648. doi: 10.1007/s10257-017-0343-1

Gupta, S., Modgil, S., Bhattacharyya, S., & Bose, I. (2021). Artificial intelligence for decision support systems in the field of operations research: review and future scope of research. *Annals of Operations Research*. doi: 10.1007/s10479-020-03856-6

Hicks, D. J. (2018). The Safety of Autonomous Vehicles: Lessons from Philosophy of Science. *IEEE Technology and Society Magazine*, *37*(1), 62–69. doi: 10.1109/MTS.2018.2795123

Huntington-Klein, N., Arenas, A., Beam, E., Bertoni, M., Bloem, J. R., Burli, P., … Stopnitzky, Y. (2021). The influence of hidden researcher decisions in applied microeconomics. *Economic Inquiry*. doi: 10.1111/ecin.12992

Inkpen, K., Chappidi, S., Mallari, K., Nushi, B., Ramesh, D., Michelucci, P., … Quinn, G. (2023). Advancing Human-AI Complementarity: The Impact of User Expertise and Algorithmic Tuning on Joint Decision Making. *ACM Transactions on Computer-Human Interaction*, *30*(5), 1–29. doi: 10.1145/3534561

Kent, D. M., Steyerberg, E., & van Klaveren, D. (2018). Personalized evidence based medicine: predictive approaches to heterogeneous treatment effects. *BMJ (Clinical Research Ed.)*, *363*, k4245. doi: 10.1136/bmj.k4245

Kleinberg, J., Lakkaraju, H., Leskovec, J., Ludwig, J., & Mullainathan, S. (2018). Human decisions and machine predictions. *The Quarterly Journal of Economics*, *133*(1), 237–293. doi: 10.1093/qje/qjx032

Kleinberg, J., Ludwig, J., Mullainathan, S., & Rambachan, A. (2018). Algorithmic Fairness. *AEA Papers and Proceedings*, *108*, 22–27. doi: 10.1257/pandp.20181018

Kliegr, T., Bahník, Š., & Fürnkranz, J. (2021). A review of possible effects of cognitive biases on interpretation of rule-based machine learning models. *Artificial Intelligence*, *295*, 103458. doi: 10.1016/j.artint.2021.103458

Kordzadeh, N., & Ghasemaghaei, M. (2022). Algorithmic bias: review, synthesis, and future research directions. *European Journal of Information Systems*, *31*(3), 388–409. doi: 10.1080/0960085X.2021.1927212

Kouri, A., Yamada, J., Lam Shin Cheung, J., Van de Velde, S., & Gupta, S. (2022). Do providers use computerized clinical decision support systems? A systematic review and meta-regression of clinical decision support uptake. *Implementation Science*, *17*(1), 21. doi: 10.1186/s13012-022-01199-3

Kuziemski, M., & Misuraca, G. (2020). AI governance in the public sector: Three tales from the frontiers of automated decision-making in democratic settings. *Telecommunications Policy*, *44*(6), 101976. doi: 10.1016/j.telpol.2020.101976

Lång, K., Josefsson, V., Larsson, A.-M., Larsson, S., Högberg, C., Sartor, H., … Rosso, A. (2023). Artificial intelligence-supported screen reading versus standard double reading in the Mammography Screening with Artificial Intelligence trial (MASAI): a clinical safety analysis of a randomised, controlled, non-inferiority, single-blinded, screening accuracy study. *The Lancet Oncology*, *24*(8), 936–944. doi: 10.1016/S1470-2045(23)00298-X

Lazer, D., Kennedy, R., King, G., & Vespignani, A. (2014). Big data. The parable of Google Flu: traps in big data analysis. *Science*, *343*(6176), 1203–1205. doi: 10.1126/science.1248506

Lehner, P. E., Adelman, L., Cheikes, B. A., & Brown, M. J. (2008). Confirmation bias in complex analyses. *IEEE Transactions on Systems, Man, and Cybernetics - Part A: Systems and Humans*, *38*(3), 584–592. doi: 10.1109/TSMCA.2008.918634

Liu, P., Du, Y., & Xu, Z. (2019). Machines versus humans: People's biased responses to traffic accidents involving self-driving vehicles. Accident Analysis and Prevention, 125, 232–240. doi: 10.1016/j.aap.2019.02.012

Logg, J. M., Minson, J. A., & Moore, D. A. (2019). Algorithm appreciation: People prefer algorithmic to human judgment. *Organizational Behavior and Human Decision Processes*, *151*, 90–103. doi: 10.1016/j.obhdp.2018.12.005

Lyell, D., & Coiera, E. (2017). Automation bias and verification complexity: a systematic review. *Journal of the American Medical Informatics Association*, *24*(2), 423–431. doi: 10.1093/jamia/ocw105

McAfee, A., & Brynjolfsson, E. (2012). Big data: the management revolution. *Harvard Business Review*, *90*(10), 60–66, 68, 128.

McKinney, S. M., Sieniek, M., Godbole, V., Godwin, J., Antropova, N., Ashrafian, H., … Shetty, S. (2020). International evaluation of an AI system for breast cancer screening. *Nature*, *577*(7788), 89–94. doi: 10.1038/s41586-019-1799-6

Mehrabi, N., Morstatter, F., Saxena, N., Lerman, K., & Galstyan, A. (2021). A survey on bias and fairness in machine learning. *ACM Computing Surveys*, *54*(6), 1–35. doi: 10.1145/3457607

Meyer, Joachim, & Kuchar, J. K. (2021). Maximal benefits and possible detrimental effects of binary decision aids. *2021 IEEE 2nd International Conference on Human-Machine Systems (ICHMS)*, 1–6. IEEE. doi: 10.1109/ICHMS53169.2021.9582632

Meyer, Joachim, & Lee, J. D. (2013). *Trust, reliance, and compliance*. Oxford University Press. doi: 10.1093/oxfordhb/9780199757183.013.0007

Meyer, Joachim. (2024). On the need to understand human behavior to do analytics of behavior. In J. Glückler & R. Panitz (Eds.), *Knowledge and digital technology* (pp. 47–62). Cham: Springer Nature Switzerland. doi: 10.1007/978-3-031-39101-9_3

Meyer, J. (2001). Effects of warning validity and proximity on responses to warnings. *Human Factors*, *43*(4), 563–572. doi: 10.1518/001872001775870395

Min, B. H., & Borch, C. (2022). Systemic failures and organizational risk management in algorithmic trading: Normal accidents and high reliability in financial markets. *Social Studies of Science*, *52*(2), 277–302. doi: 10.1177/03063127211048515

Mosier, K. L., Skitka, L. J., Heers, S., & Burdick, M. (1997). Automation bias: decision making and performance in high-tech cockpits. *The International Journal of Aviation Psychology*, *8*(1), 47–63. doi: 10.1207/s15327108ijap0801_3

Ng, K. K. H., Chen, C.-H., Lee, C. K. M., Jiao, J. (Roger), & Yang, Z.-X. (2021). A systematic literature review on intelligent automation: Aligning concepts from theory, practice, and future perspectives. *Advanced Engineering Informatics*, *47*, 101246. doi: 10.1016/j.aei.2021.101246

Padilla, R., Netto, S. L., & da Silva, E. A. B. (2020). A Survey on Performance Metrics for Object-Detection Algorithms. *2020 International Conference on Systems, Signals and Image Processing (IWSSIP)*, 237–242. IEEE. doi: 10.1109/IWSSIP48289.2020.9145130

Parasuraman, R, Sheridan, T. B., & Wickens, C. D. (2000). A model for types and levels of human interaction with automation. *IEEE Transactions on Systems, Man, and Cybernetics - Part A: Systems and Humans*, *30*(3), 286–297. doi: 10.1109/3468.844354

Parasuraman, Raja, & Riley, V. (1997). Humans and automation: use, misuse, disuse, abuse. *Human Factors*, *39*(2), 230–253. doi: 10.1518/001872097778543886

Power, D. J. (2008). Understanding Data-Driven Decision Support Systems. *Information Systems Management*, *25*(2), 149–154. doi: 10.1080/10580530801941124

Price, H. E. (1985). The allocation of functions in systems. *Human Factors*, *27*(1), 33–45. doi: 10.1177/001872088502700104

Puaschunder, J. M., Mantl, J., & Plank, B. (2020). Medicine of the future: The power of Artificial Intelligence (AI) and big data in healthcare. *Zenodo*. doi: 10.5281/zenodo.3839002

Richardson, S. (2022). Exposing the many biases in machine learning. *Business Information Review*, *39*(3), 82–89. doi: 10.1177/02663821221121024

Roig, A. (2017). Safeguards for the right not to be subject to a decision based solely on automated processing (Article 22 GDPR). *European Journal of Law and Technology*.

Roth, E. M., Sushereba, C., Militello, L. G., Diiulio, J., & Ernst, K. (2019). Function allocation considerations in the era of human autonomy teaming. *Journal of Cognitive Engineering and Decision Making*, *13*(4), 199–220. doi: 10.1177/1555343419878038

Saxena, D., Badillo-Urquiola, K., Wisniewski, P. J., & Guha, S. (2020). A Human-Centered Review of Algorithms used within the U.S. Child Welfare System. *Proceedings of the 2020 CHI Conference on Human Factors in Computing Systems*, 1–15. New York, NY, USA: ACM. doi: 10.1145/3313831.3376229

Schweinsberg, M., Feldman, M., Staub, N., van den Akker, O. R., van Aert, R. C. M., van Assen, M. A. L. M., … Luis Uhlmann, E. (2021). Same data, different conclusions: Radical dispersion in empirical results when independent analysts operationalize and test the same hypothesis. *Organizational Behavior and Human Decision Processes*, *165*, 228–249. doi: 10.1016/j.obhdp.2021.02.003

Sidi, F., Shariat Panahy, P. H., Affendey, L. S., Jabar, M. A., Ibrahim, H., & Mustapha, A. (2012). Data quality: A survey of data quality dimensions. *2012 International Conference on Information Retrieval & Knowledge Management*, 300–304. IEEE. doi: 10.1109/InfRKM.2012.6204995

Silberzahn, R., Uhlmann, E. L., Martin, D. P., Anselmi, P., Aust, F., Awtrey, E., … Nosek, B. A. (2018). *Many analysts, one data set: making transparent how variations in analytic choices affect results.* *1*(3), 337–356. doi: 10.1177/2515245917747646

Susanto, A. P., Lyell, D., Widyantoro, B., Berkovsky, S., & Magrabi, F. (2023). Effects of machine learning-based clinical decision support systems on decision-making, care delivery, and patient outcomes: a scoping review. *Journal of the American Medical Informatics Association*, *30*(12), 2050–2063. doi: 10.1093/jamia/ocad180

Sutton, R. T., Pincock, D., Baumgart, D. C., Sadowski, D. C., Fedorak, R. N., & Kroeker, K. I. (2020). An overview of clinical decision support systems: Benefits, risks, and strategies for success. *Npj Digital Medicine*, *3*(1), 17. doi: 10.1038/s41746-020-0221-y

Takemura, K. (2021). *Behavioral decision theory: psychological and mathematical descriptions of human choice behavior*. Singapore: Springer Singapore. doi: 10.1007/978-981-16-5453-4

Tao, R., Su, C.-W., Xiao, Y., Dai, K., & Khalid, F. (2021). Robo advisors, algorithmic trading and investment management: Wonders of fourth industrial revolution in financial markets. *Technological Forecasting and Social Change*, *163*, 120421. doi: 10.1016/j.techfore.2020.120421

Veale, M., & Zuiderveen Borgesius, F. (2021). Demystifying the Draft EU Artificial Intelligence Act — Analysing the good, the bad, and the unclear elements of the proposed approach. *Computer Law Review International*, *22*(4), 97–112. doi: 10.9785/cri-2021-220402

Virgilio, G. P. M. (2019). High-frequency trading: a literature review. *Financial Markets and Portfolio Management*, *33*(2), 183–208. doi: 10.1007/s11408-019-00331-6

Wexler, J., Pushkarna, M., Bolukbasi, T., Wattenberg, M., Viegas, F., & Wilson, J. (2020). The What-If Tool: Interactive Probing of Machine Learning Models. *IEEE Transactions on Visualization and Computer Graphics*, *26*(1), 56–65. doi: 10.1109/TVCG.2019.2934619

White, N. M., Carter, H. E., Kularatna, S., Borg, D. N., Brain, D. C., Tariq, A., … McPhail, S. M. (2023). Evaluating the costs and consequences of computerized clinical decision support systems in hospitals: a scoping review and recommendations for future practice. *Journal of the American Medical Informatics Association*, *30*(6), 1205–1218. doi: 10.1093/jamia/ocad040

Wickens, C. D., Clegg, B. A., Vieane, A. Z., & Sebok, A. L. (2015). Complacency and automation bias in the use of imperfect automation. *Human Factors*, *57*(5), 728–739. doi: 10.1177/0018720815581940

Zarsky, T. (2016). The Trouble with Algorithmic Decisions. *Science, Technology, & Human Values*, *41*(1), 118–132. doi: 10.1177/0162243915605575